\documentclass[twocolumn,showpacs,preprintnumbers,amsmath,amssymb]{revtex4}
\usepackage{graphicx}
\usepackage{dcolumn}
\usepackage{bm}

\def\ep{\varepsilon}

\def\Lm{\Lambda}

\begin{document}
\title{Localization-delocalization transition in one-dimensional electron systems
with long-range correlated disorder}
\author{H.~Shima, T.~Nomura and T.~Nakayama}
\affiliation{Department of Applied Physics, 
Graduate School of Engineering,
Hokkaido University, Sapporo 060-8628, Japan}
\date{\today}

%
%

\begin{abstract}
We investigate localization properties of electron eigenstates
in one-dimensional (1d) systems with long-range correlated diagonal disorder.
Numerical studies on the localization length $\xi$ of eigenstates
demonstrate the existence of the localization-delocalization transition in 1d systems
and elucidate non-trivial behavior of $\xi$ as a function of the disorder strength.
The critical exponent $\nu$ for localization length is extracted
for various values of parameters characterizing the disorder,
revealing that every $\nu$ disobeys the Harris criterion $\nu > 2/d$.
\end{abstract} 

%

\maketitle

\section{Introduction}

Spatial correlation of disorder often causes an unexpected phenomenon
in quantum disordered systems.
Most intriguing is the breakdown of Anderson localization
in one-dimensional (1d) systems induced by correlated disorder.
Breakdown of the localization in 1d systems
has been predicted for the first time for a random dimer model \cite{dimer1},
wherein the on-site potential $\{\ep_i\}$
has a binary distribution with a {\it short-range} spatial correlation.
Subsequently, a discrete number of extended eigenstates was found numerically 
in the random dimer model \cite{dimer2,dimer3}. It was
examined experimentally using transmission measurements 
on semiconductor superlattices \cite{Bellani}.
These findings have motivated the studies of the nature
of 1d systems with {\it long-range} correlated disorder \cite{group13,14,15,16,17,18,19,20}.
Particularly noteworthy is the system in which the sequence $\{\ep_i\}$
has a power-law spectral density of the form $S(k)\propto k^{-p}$:
$S(k)$ is the Fourier transform of the spatial correlation function
$\langle \ep_i \ep_j \rangle$.
For exponents $p$ greater than $2.0$,
there is a finite range of energy values with extended eigenstates \cite{14}.
This fact indicates the presence of the localization-delocalization transition
in 1d disordered systems
against the conclusion of the well-known scaling theory
\cite{Anderson, Scaling}.
The emergence of extended eigenstates was also observed in harmonic chains 
with random couplings \cite{Adame} and in those with randoms masses \cite{Moura3}.
Note that diagonal disorder treated in Ref.~\cite{Moura3} is characterized by 
the power-law spectral density denoted above.

The randomness of the long-range correlated potentials $\{\ep_i\}$
is characterized by two quantities.
The first is the exponent of the power-law spectral density, $p$,
determining the roughness of potential landscapes, as shown in Fig.~1.
The second quantity is the distribution width $W$ defined by the relation $\ep_i \in [-W/2, W/2]$,
which characterizes the amplitude of the potential \cite{foot}.
Effects of $p$ on the localization properties of eigenstates
have been examined \cite{14}, but
those of $W$ remain unclarified.
If the disorder is spatially {\it uncorrelated} ($p=0$),
an increase in $W$ trivially induces strongly-localized wavefunctions 
because all eigenstates are localized exponentially 
\cite{Ishii,Scaling}.
On the other hand, when the disorder is sufficiently long-range correlated
to yield extended eigenstates ($p>2.0$),
the system shows a critical point $W_c$
separating localized and delocalized phases.
However, there is no attempt to quantitatively determine the value of $W_c$.
Calculations for $W_c$ for various values of $p$ allow us 
to establish the phase diagram in the $W$-$p$ space,
thereby engendering better understanding of localization properties of the system.
We are also interested in the value of the critical exponent $\nu$
for the localization length of eigenstates.
Those values can be obtained accurately using finite-size scaling analysis.

The present work is intended to reveal critical properties of electron eigenstates
in 1d systems with long-range correlated disorder.
Numerical studies on the localization lengths $\xi$
have demonstrated the non-trivial behavior of $\xi$ as a function of $W$ and $p$.
A series of critical widths $W_c$ and that of critical exponents $\nu$
are determined by finite-size scaling analysis.
Remarkably, the results of $\nu$ disobey the Harris criterion, $\nu > 2/d$
\cite{Harris, Chayes},
which is believed to be satisfied in general disordered systems.
Our findings introduce new prospects 
for the study of Anderson transition in 1d systems with correlated disorder.

This paper is organized as follows.
Section II describes the long-range correlated on-site potentials to be considered.
It presents a numerical algorithm for calculating the localization length of eigenstates.
Section III analyzes the localization length as a function of $W$ and $p$.
The transition point $W_c$ and the critical exponent $\nu$ are extracted
by finite-size scaling analysis.
Conclusions and discussion are presented in Sec. IV.

\section{Model and Method}

\subsection{Long-range correlated potentials}

We consider noninteracting electrons in 1d disordered systems
within a tight binding approximation.
The Schr\"odinger equation of the system is expressed as
\begin{equation}
\ep_i \phi_i + t(\phi_{i+1}+\phi_{i-1})=E\phi_i,
\label{eq01}
\end{equation}
where $\phi_i$ is the amplitude of the wavefunction at the $i$-th site of the lattice.
The hopping energy $t$ is taken as a unit of energy hereafter.
A sequence of long-range correlated potential $\{\ep_i\}$
is produced by the Fourier filtering method \cite{FFM2,FFM4,FFM5}.
This method is based on a transformation of the Fourier components 
of a random number sequence.
The outline of the method is:
i) A sequence $\{u_i\}$ of uncorrelated random numbers 
with a Gaussian distribution is prepared.
ii) Its Fourier components $\{u_q\}$ are computed
using the fast-Fourier transformation method.
iii) A sequence $\{\ep_q \}$ is newly generated for a given $p$ using the relation $\ep_q = q^{-p/2} u_q$.
iv) Finally, the objective $\{ \ep_i\}$ is obtained
as the inverse Fourier transform of $\{\ep_q\}$.
The resulting potentials $\ep_i$ are spatially correlated and
produce the power-law spectral density
$S(k) \propto k^{-p}$.
In the following, the mean value $\langle \ep_i \rangle$ is set to be zero
and the periodic boundary condition is imposed.

\begin{figure}[ttt]
\hspace*{-0.5cm}
\includegraphics[width=7.8cm]{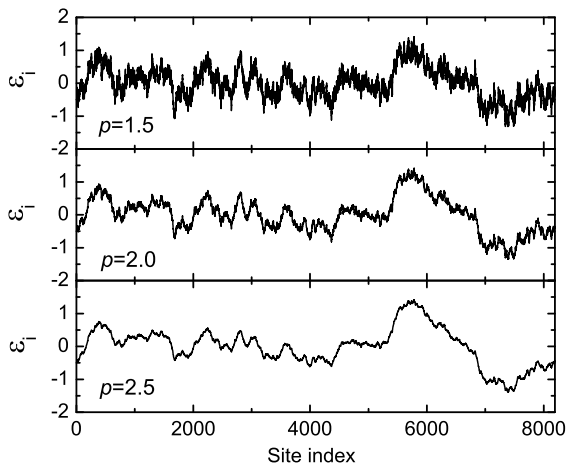}
\caption{Landscapes of spatially correlated on-site potentials.
The exponent $p$ of the power-law spectral density 
$S(k) \propto k^{-p}$ is varied as displayed in the figure.
The roughness of the energy landscape is gradually reduced with increasing $p$.}
\label{fig1}
\end{figure}

Figure 1 displays the landscapes of long-range correlated potentials $\ep_i$
that are generated by the procedure above.
The system size $L=2^{13}$ and the distribution width $W=3.0$
are fixed.
An increase in $p$ markedly reduces the roughness
in potential landscapes.
We have confirmed that the sequences $\{\ep_i\}$ appearing in Fig.~1
produce the power-law spectral density,
$S(k) \propto k^{-p}$, for all values of $p$.

\subsection{Localization lengths}

Localization lengths of eigenstates in a potential field $\{\ep_i\}$
are computed easily using the conventional transfer matrix method \cite{Kramer}.
The Schr\"odinger equation (\ref{eq01}) is expressed by the following matrix equation:
\begin{equation}
\left(
\begin{array}{c}
\phi_{i+1}\\
\phi_i
\end{array}
\right)=
\bm{M}_i
\left(
\begin{array}{c}
\phi_i\\
\phi_{i-1}
\end{array}
\right)
,
\quad
\bm{M}_i \equiv \left(
\begin{array}{cc}
E-\ep_i & -1\\
1 & 0
\end{array}
\right).
\label{eq02}
\end{equation}
The localization length $\xi$ at a given energy $E$ is defined by the relation \cite{Kramer}
\begin{equation}
\xi^{-1} = \lim_{L\to \infty} \frac1L \ln \frac{\left| \Pi_{i=1}^N \bm{M}_i z(0) \right|}
{\left| z(0)\right|}
\label{eq03}
\end{equation}
with a generic initial condition 
\begin{equation}
z(0)= \left(
\begin{array}{c}
\phi_1 \\
\phi_0
\end{array}
\right).
\label{eq04}
\end{equation}
Equation (\ref{eq03}) gives the single value of $\xi$
only for the infinite system size $L\to \infty$.
However, when $L$ is finite, the calculated result of Eq.~(\ref{eq03})
-- denoted as $\xi_L$ --
depends on the choice of the potential field $\{\ep_i\}$.
To obtain a typical value of $\xi_L$ for a given $L$,
we take a geometrical mean of $\xi_L$ on more than $10^4$ samples.
Energy $E$ is fixed at the band center $E=0$ throughout this paper.

A critical point $W=W_c$ (and $p=p_c$)
can be deduced from the dependence of the normalized localization length 
$\Lm\equiv \xi_L/L$ on the system size $L$.
The typical values of $\xi_L$ increase with $L$ for delocalized states,
where the growth of $\xi_L$ is faster than that in $L$.
This causes the quantity $\Lm$ to be an increasing function of $L$.
On the other hand, $\Lm$ vanishes for sufficiently large $L$ for localized states,
because $\xi_L$ approaches a constant value.
Therefore, at the localization-delocalization transition, 
$\Lm$ must be invariant for the change in the system size $L$.
Values of $W_c$ are obtained accurately using the finite-size scaling method,
as explained in the next section.

\section{Numerical Results}

\begin{figure}[ttt]
\includegraphics[width=7.5cm]{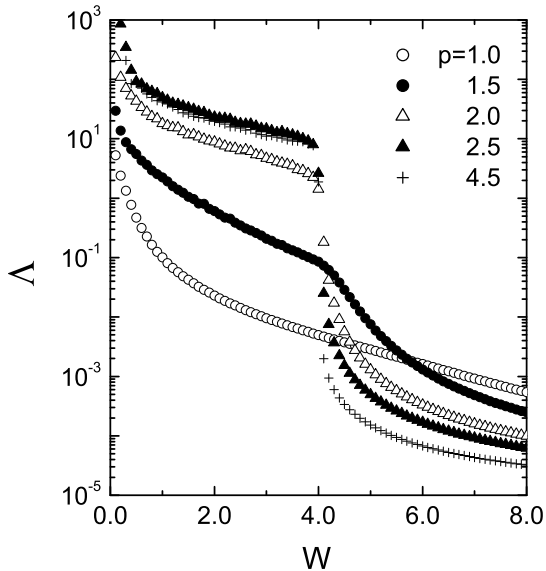}
\caption{The normalized localization length $\Lm\equiv \xi_L/L$
as a function of the distribution width $W$.
Several values of $p$ are taken as denoted in the figure.
The system size $L$ is fixed to be $2^{16}$ for all values of $p$.
}
\label{fig2}
\end{figure}

\subsection{The $W$-dependence for the function $\Lm(W)$}

Figure 2 plots the normalized localization length $\Lm$
as a function of the distribution width $W$.
The exponent $p$ of the power-law spectral density $S(k)$
is increased from $p=1.0$ to $p=4.5$ incrementally.
The system size $L=2^{16}$ is fixed for all $p$.
When $p$ equals unity or less, $\Lm(W)$ is a monotonous function
of $W$.
For larger $p$, on the other hand, curves of $\Lm$ show a kink
at $W=4.0$, which sharpens as $p$ increases.

We find two striking features in Fig.~2.
The first is a peculiar $p$-dependence of $\Lm(W)$.
The $p$-dependence of $\Lm(W)$ in the region where $W<4.0$
differs completely from that in the region where $W>4.0$.
For $W<4.0$, the values of $\Lm$ rise with increasing $p$ indicating that the growth of $p$ for $W<4.0$ causes an increase 
in the localization length $\xi_L$ of eigenstates.
In contrast, the values of $\Lm$ for $W>4.0$ systematically decrease 
with increasing $p$ (except for the data $p=1.0$).
Hence, growth of $p$ for $W>4.0$ produces strongly-localized eigenstates.
This difference in the effect of increasing $p$ on $\xi_L$ is non-trivial
because the increase in $p$ simply smoothes potential landscapes,
as shown in Fig.~1.
Therefore our results suggest that 
the effect of potential roughness on the localization length $\xi_L$
depends strongly on the value of $W$.

The second notable feature is a shoulder structure of $\Lm(W)$
-- a sharp bent of $\Lm(W)$ -- at around $W=4.0$,
engendering a plateau-like shape within the region $0.5<W<4.0$.
The shoulder of $\Lm(W)$ appears for exponents $p>2.0$,
i.e., for $p$ large enough to yield extended eigenstates \cite{14}.
That fact implies that localization properties of eigenstates for $W<4.0$
(and $p>2.0$) differ substantially from those for $W>4.0$.
The value $W=4.0$ is a critical disorder strength
separating a localized and delocalized phase, as explained later.
The next subsection presents a demonstration that eigenstates are delocalized within the plateau region,
under conditions $W<4.0$ and $p>2.0$.

\begin{figure}[ttt]
\includegraphics[width=7.8cm]{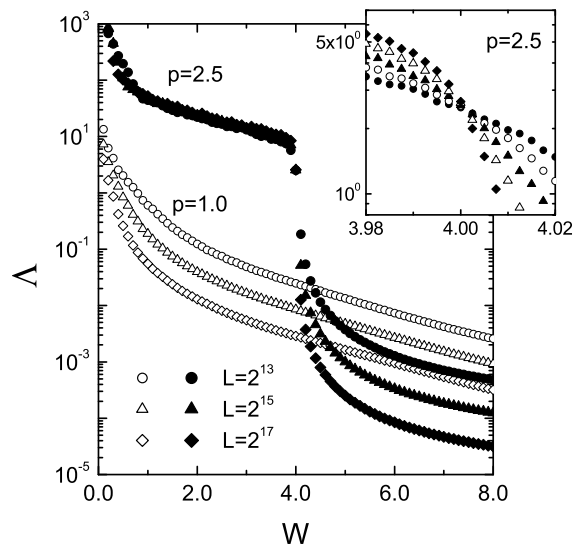}
\caption{The $L$-dependence for $\Lm(W)$
with setting $p=1.0$ (open symbols) and $p=2.5$ (solid symbols).
Inset: Enlargement of $\Lm(W)$ in the vicinity of $W=4.0$.
The system size $L$ is varied to $L=2^{13}$ (solid circles), $2^{14}$ (open circles),
$2^{15}$ (solid triangles), $2^{16}$ (open triangles), and $2^{17}$ (diamonds).
}
\label{fig3}
\end{figure}

Values of the critical width $W_c$ can be estimated from
the dependence of $\Lm(W)$ on system size $L$.
Figure 3 shows the $L$-dependence for $\Lm(W)$,
where the system size is increased from $L=2^{13}$ (circles) up to $2^{17}$ (diamonds).
Open and solid symbols correspond to the exponent $p=1.0$ and $p=2.5$, respectively.
For $p=1.0$, the magnitude of $\Lm(W)$ declines with increasing $L$ such that 
the eigenstates are localized for any $W$.
On the other hand, for $p=2.5$, the $L$-dependence for $\Lm(W)$ is rather complicated.
The inset of Fig.~3 displays detailed behavior of $\Lm$ in the vicinity of $W=4.0$,
where the system size is varied from $L=2^{13}$ to $2^{17}$ as denoted 
in the figure caption.
All curves intersect on a single point
at $W=4.0$, suggesting the presence of the Anderson transition
at $W=W_c = 4.0$ for $p=2.5$.

We have confirmed that, when $p\ge 2.0$, $\Lm(W)$ shows the same $L$-dependence
as presented in the inset of Fig.~3.
Intriguingly, all data of $\Lm(W)$ for $p>2.0$ provide an identical value of $W_c=4.0$.
Therefore, we conclude that the critical distribution width $W_c=4.0$
is independent of $p$ whenever $p \ge 2.0$.

\subsection{Finite-size scaling analysis}

Finite-size scaling analysis allows the determination of critical properties 
of the transition for $L\to \infty$ from data for finite $L$ \cite{Kramer}.
This method stems from the hypothesis that the normalized localization length $\Lm$
close to the transition obeys the scaling law expressed by
\begin{equation}
\ln \Lm = f\left(\frac{L}{\xi_{\infty}}\right),
\label{eq05}
\end{equation}
where $\xi_{\infty}$ is the localization length of eigenstates
in an infinite system.
The argument $L/\xi_{\infty}$
becomes much smaller than unity because $\xi_{\infty}$ diverges with obeying the form
$\xi_{\infty} \propto |W-W_c|^{-\nu}$ near the transition point $W=W_c$.
This allows expansion of the scaling function as
\begin{equation}
\ln \Lm = a_0 + a_1 |W-W_c|L^{1/\nu} + \cdots + a_{n} |W-W_c|^n L^{n/\nu},
\label{eq06}
\end{equation}
terminating the expansion at the order $n$.
Fitting the numerical data of $\ln \Lm$ for various values of $W$ and $L$
to Eq.~(\ref{eq06}),
we obtain $W_c$ and $\nu$ with great accuracy.
Note that the optimal value of constants
$a_0$, $a_1$, $\cdots$, $a_n$ for $W>W_c$ are different from those for $W<W_c$
because $\xi_{\infty}$ is a function of the absolute value of $W-W_c$.

\begin{figure}[ttt]
\includegraphics[width=7.5cm]{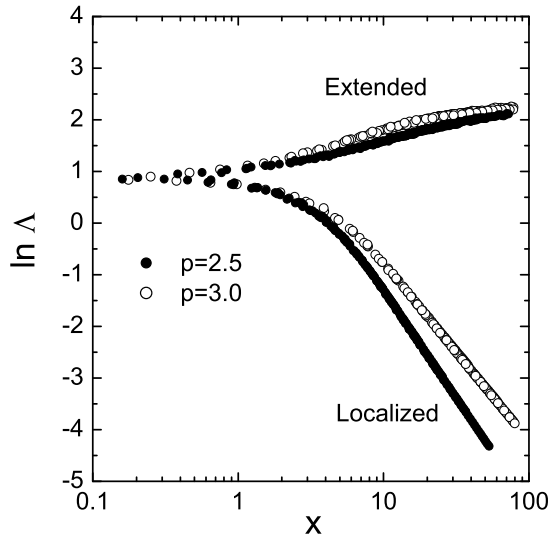}
\caption{Scaling plots of $\ln \Lm=f(x)$ for $p=2.5$
(solid circles) and for $p=3.0$ (open circles).
Upper and lower branches correspond to 
extended and localized phases, respectively.
}
\label{fig4}
\end{figure}


\begin{table}[bbb]
\caption{Calculated results of the critical exponent $\nu$ 
and the critical distribution width $W_c$.
The error is a 95\% confidence interval.}
\begin{ruledtabular}
\begin{tabular}{cccc}
 & Branch & $\nu$ & $W_c$ \\

\hline
 $p=2.5$ & Extended & 1.79$\pm$0.08 & 4.002$\pm$0.002 \\
         & Localized & 1.87$\pm$0.04 & 4.001$\pm$0.001 \\
\hline
 $p=3.0$ & Extended & 1.58$\pm$0.08 & 4.001$\pm$0.002 \\
         & Localized & 1.51$\pm$0.02 & 4.001$\pm$0.001
         \end{tabular}
\end{ruledtabular}
\end{table}

Figure 4 shows scaling plots of $\ln \Lm$ for $p=2.5$ (solid) 
and $p=3.0$ (open).
Here we define $x\equiv |W-W_c|L^{1/\nu}$ and set $n=4$.
Each upper and lower branch in the figure corresponds to the extended phase ($W<W_c$)
and the localized phase ($W>W_c$), respectively.
All data of $\ln \Lm$ for various values of $W$ and $L$ 
fit well onto two branches.
Resulting values of the critical exponent $\nu$ 
and the critical distribution width $W_c$ are listed in Table~1,
where the error is a 95\% confidence interval.
We see that all values of $W_c$ are almost identical to $4.0$, as expected.
On the other hand, 
values of $\nu$ exhibit a discrepancy for different branches
and different $p$s \cite{foot1}.
This discrepancy of $\nu$ contradicts the principle of one-parameter scaling
requiring that $\nu$ should be independent of a choice of parameters
in the Hamiltonian of the system.
It is surmised that the discrepancy of $\nu$ occurs because of a finite-sized effect
that causes a systematic error in scaling plots.
A novel technique of scaling correction (two-parameter scaling)
\cite{two-para1, two-para2, two-para3}
would help to solve the problem,
as we shall present in a future study \cite{future}.

\begin{figure}[ttt]
\includegraphics[width=7.4cm]{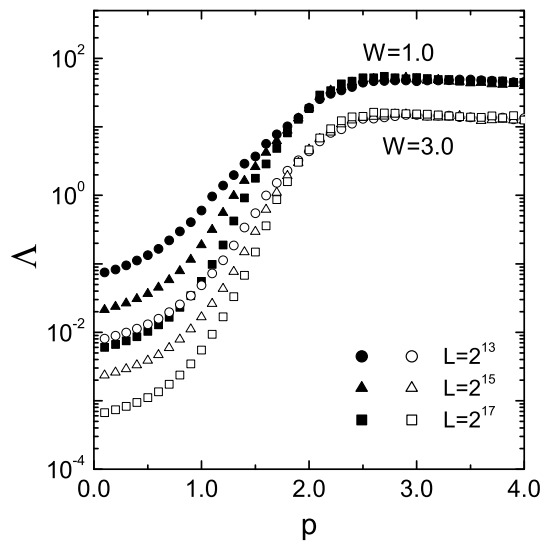}
\caption{The $L$-dependence for $\Lm(p)$
with setting $W=1.0$ (solid symbols) and $W=3.0$ (open symbols), respectively.
The system size $L$ is varied as $2^{13}$ (circle), $2^{15}$ (triangle), and 
$2^{17}$ (rectangular).
}
\label{fig5}
\end{figure}

\subsection{The $p$-dependence for the function $\Lm(p)$}

Next, we examine $\Lm$ behavior as a function of $p$.
Figure 5 shows a plot of the function $\Lm(p)$ with setting $W=1.0$ and $3.0$,
where the system size is varied from $L=2^{13}$ to $2^{17}$.
For all $W$ and $L$,
$\Lm(p)$ increases with $p$; 
it then becomes constant for $p>2.5$.
This $p$-dependence for $\Lm$ is consistent with the results displayed in Fig.~2,
in which the growth of $p$ for $W<4.0$ causes an increase in $\Lm$.
Detail calculations of $\Lm(p)$ in the vicinity of $p=2.0$
reveal that all curves of $\Lm(p)$ that belong to different $L$s intersect
at the point at $p=2.0$ as long as $W<4.0$.
This indicates that localization-delocalization transition occurs 
at an identical point $p=p_c=2.0$ whenever $W<4.0$.
The critical point $p_c=2.0$ estimated above
is consistent with the conclusion reported in Ref.~\cite{14}.

A series of critical points $W_c$ and $p_c$ we have found
are summarized in the phase diagram
illustrated in Fig.~6.
Solid circles express the critical point $W=W_c$ 
deduced from the finite-size scaling procedure,
whereas solid rectangulars express $p=p_c$ defined by the position
at which $\Lm(p)$ is independent of $L$ (See Fig.~5).
An extended phase appears in the region surrounded by the two straight lines of
$W=4$ and $p=2$.
It is noteworthy that values of $W_c$ 
and $p_c$ equal an {\it integral} number: $4$ and $2$, respectively.
The same integral value of $W_c=4$
has been observed in other 1d systems
with correlated disorder \cite{dimer1,dimer2,dimer3,Harper,Hof,Aubry},
as discussed below.

\begin{figure}[ttt]
\includegraphics[width=6cm]{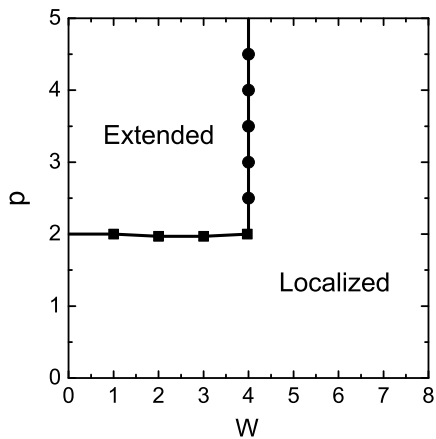}
\caption{A phase diagram of the system in the $W$-$p$ space.
The critical line separating the localized and delocalized phase
consists of two straight lines: $W=W_c=4.0$ and $p=p_c=2.0$.}
\label{fig7}
\end{figure}

\section{Concluding remarks}

Two open problems remain with respect to the critical properties of the 
transition in 1d systems with long-range correlated disorder.
First, all results of the critical exponent $\nu$
listed in Table 1 disobey the Harris criterion $\nu>2/d$ \cite{Harris, Chayes}.
The inequality is widely believed to be satisfied
in general disordered systems with any spatial dimension $d$,
thereby determining the lower bound for $\nu$.
According to the inequality, $\nu$ in 1d systems must be larger than $2$,
which disagrees with our results.
We note here that the Harris criterion 
was originally derived for a spatially {\it uncorrelated} disorder 
\cite{Harris}.
Therefore, the relation $\nu>2/d$ may be violated in systems
with long-range {\it correlated} disorder.
In fact, for {\it classical} percolation model,
the inequality must be modified in the presence of long-range correlation
in the site or bond occupations
\cite{Weinrib1, Weinrib2}.
To elucidate the lower bound of $\nu$ for 1d systems considered,
we should generalize the argument in Ref.~\cite{Chayes}
for the correlated disorder producing the power-law spectral density
$S(k)\propto k^{-p}$.

Secondly, the critical disorder width
$W_c$ is exactly equal to an {\it integral} number $W_c=4$.
Furthermore, the value of $W_c$
is independent of the potential roughness characterized by $p$.
In fact, the same value of $W_c=4$ has been found
in other correlated potential models.
One example is the random dimer model \cite{dimer1,dimer2,dimer3},
where the site energies $\ep_a$ and $\ep_b$ 
are assigned in pairs.
The disorder width $W$ of the system is defined 
by the relation $\ep_a-\ep_b \in [-W/2, W/2]$.
If $W$ is less than the critical value $W_c=4$,
the localization length diverges obeying the form $\xi_{\infty}(E) \propto E^{-2}$,
yielding the delocalized eigenstate at $E=0$.
Another example is the Harper model \cite{Harper,Hof},
in which site energies are described by a periodic function
$\ep_i=(W/2) \cos(2\pi i \omega)$ with an irrational number $\omega$.
This model undergoes the localization-delocalization transition
at $W=4$. That is, all eigenstates become critical at $W=4$
\cite{Aubry}.
We conjecture that the coincidence of the value of $W_c$ in the three different models
is evidence of an unexplored universality of $W_c$
in 1d systems with spatially correlated disorder \cite{foot2}.

In conclusion, we have investigated critical properties of
1d electron systems subject to long-range correlated disorder.
The normalized localization length $\Lm$ of eigenstates
shows a non-trivial behavior as a function of $W$ and $p$,
which indicates a peculiarity of 1d systems
exhibiting the Anderson transition.
Detailed calculations for $\Lm(W,p)$ reveal that the transition point $W_c=4$
is invariant to the change of the potential roughness.
Moreover, the critical exponent $\nu$ in the considered system 
is proven to violate the Harris criterion $\nu > 2/d$.
We hope that our findings enlighten the study of 
quantum phase transition in manifold disordered systems.

\begin{acknowledgments}

We are grateful to K.~Yakubo for helpful discussion.
This work is supported financially by a Grant-in-Aid for Scientific
Research from the Japan Ministry of Education, Science, Sports and Culture.
Numerical calculations were performed in part on the SR8000
of the Supercomputer Center, ISSP, University of Tokyo.

\end{acknowledgments}

%
%


\begin{thebibliography}{99}

\bibitem{dimer1} D.~H.~Dunlap, H-~L.~Wu, 
and P.~W.~Phillips, Phys.~Rev.~Lett. {\bf 65}, 88 (1990).


\bibitem{dimer2} S.~N.~Evangelou and A.~Z.~Wang, Phys.~Rev.~B {\bf 47}, 13126 (1993)


\bibitem{dimer3} S.~N.~Evangelou and E.~N.~Economou,
J.~Phys.~A: Math.~Gen. {\bf 26}, 2803 (1993).


\bibitem{Bellani} V.~Bellani, E.~Diez, R.~Hey, L.~Toni, L.~Tarricone,
G.~B.~Parravicini, F.~Dom\'{\i}nguez-Adame, and R.~Gom\'ez-Alcal'a,
Phys.~Rev.~Lett. {\bf 82}, 2159 (1999).



\bibitem{group13} H.~Yamada, M.~Goda, and Y.~Aizawa, 
J.~Phys.~Condens.~Matter {\bf 3}, 10043 (1991);
J.~Phys.~Soc.~Jpn. {\bf 60}, 3501 (1991).


\bibitem{14} F.~A.~B.~F.~de Moura and M.~L.~Lyra, 
Phys. Rev. Lett. {\bf 81}, 3735 (1998).


\bibitem{15} F.~M.~Izrailev and A.~A.~Krokhin, 
Phys.~Rev.~Lett. {\bf 82}, 4062 (1999).


\bibitem{16} H.~Yamada and T.~Okabe, Phys.~Rev.~E {\bf 63}, 026203 (2001).


\bibitem{17} S.~Russ, J.~W.~Kantelhardt, A.~Bunde and S.~Havlin, 
Phys.~Rev.~B {\bf 64}, 134209 (2001).


\bibitem{18} L.~Tessieri and F.~M.~Izrailev, Phys.~Rev.~E {\bf 64}, 066120 (2001).


\bibitem{19} P.~Carpena, P.~B.~Galvan, P.~Ch.~Ivanov, and H.~E.~Stanley, 
Nature (London) {\bf 418}, 955 (2002)


\bibitem{20} L.~I.~Deych, M.~V.~Erementchouk, and A.~A.~Lisyansky,
Phys.~Rev.~B {\bf 67}, 024205 (2003).



\bibitem{Anderson} P.~W.~Anderson, Phys.~Rev. {\bf 109}, 1492 (1958).


\bibitem{Scaling} E.~Abrahams, P.~W.~Anderson, D.~C.~Licciardello, 
and T.~V.~Ramakrishnan, Phys.~Rev.~Lett. {\bf 42}, 673 (1979).



\bibitem{Adame} F.~Dom\'{\i}nguez-Adame, E.~Mac\'{\i}a and A.~S\'anchez,
Phys.~Rev.~B {\bf 48}, 6054 (1993).


\bibitem{Moura3} F.~A.~B.~F.~ de Moura, M.~D.~Coutinho-Filho, E.~P.~Raposo, and
M.~L.~Lyra, Phys.~Rev.~B {\bf 68}, 012202 (2003).


\bibitem{foot} It is possible to control the amplitude of on-site potentials $\ep_i$
by the variance $\sigma \equiv \sqrt{\langle\ep_i^2\rangle - \langle\ep_i\rangle^2}$
instead of $W$.
In fact, the normalization condition $\sigma=1$ has been adopted in Ref.~\cite{Moura3}
for establishing potential sequences.
The pertinence of this normalization procedure was discussed in Refs.~\cite{comment} and \cite{reply}.

\bibitem{comment} J.~W.~Kantelhardt, S.~Russ, A.~Bunde, S.~Havlin, and I.~Webman,
Phys.~Rev.~Lett. {\bf 84}, 198 (2000).

\bibitem{reply} F.~A.~B.~F.~de Moura and M.~L.~Lyra, Phys.~Rev.~Lett. {\bf 84}, 199 (2000).


\bibitem{Ishii} K.~Isii, Prog.~Theor.~Phys.~Suppl. {\bf 53}, 77 (1973).

\bibitem{Harris} A.~B.~Harris, J.~Phys.~C {\bf 7}, 1671 (1974);
Z.~Phys.~B {\bf 49}, 347 (1983).


\bibitem{Chayes} J.~T.~Chayes, L.~Chayes, D.~S.~Fisher, and T.~Spencer,
Phys.~Rev.~Lett. {\bf 57}, 2999 (1986). 


\bibitem{FFM2} D.~Saupe, 
in {\it The Science of Fractal Images},
edited by H.~-O.~Peitgen and D.~Saupe (Springer, New York, 1988);
J.~Feder, {\it Fractals} (Plenum Press, New York, 1988). 


\bibitem{FFM4} C.~-K.~Peng, S.~Havlin, M.~Schwartz and H.~E.~Stanley,
Phys.~Rev.~A {\bf 44}, 2239 (1991) . 


\bibitem{FFM5} S.~Prakash, S.~Havlin, M.~Schwartz and H.~E.~Stanley,
Phys.~Rev.~A {\bf 46}, R1724 (1992) . 


\bibitem{Kramer} B.~Kramer and  A.~MacKinnon, Rep.~Prog.~Phys. {\bf 56}, 1469 (1993).



\bibitem{foot1} We have numerically confirmed,
at least for $2.0\le p\le 4.5$, that the critical disorder width $W_c$
is independent of $p$ and the critical exponent
$\nu$ decreases monotonically with increasing $p$.



\bibitem{two-para1} B.~Huckestein, Rev.~Mod.~Phys. {\bf 67}, 357 (1995).


\bibitem{two-para2} K.~Slevin and T.~Ohtsuki, Phys.~Rev.~Lett. {\bf 82}, 382 (1999).


\bibitem{two-para3} K.~Slevin, P.~Markos, and T.~Ohtsuki,
Phys.~Rev.~Lett. {\bf 86}, 3594 (2001).


\bibitem{future} H.~Shima and T.~Nakayama, {\it to be published}.

\bibitem{Weinrib1} A.~Weinrib and B.~I.~Halperin, Phys.~Rev.~B {\bf 27}, 413 (1983).


\bibitem{Weinrib2} A.~Weinrib, Phys.~Rev.~B {\bf 29}, 387 (1984).


\bibitem{Harper} P.~G.~Harper, Proc.~Phys.~Soc.~Lond.~A {\bf 68}, 874 (1955). 


\bibitem{Hof} D.~R.~Hofstadter, Phys.~Rev.~B {\bf 14}, 2239 (1976).


\bibitem{Aubry} A.~Aubry and G.~Andr\'e, Ann.~Isr.~Phys.~Soc. {\bf 3}, 133 (1980).

\bibitem{foot2}
It might be the case that the supposed universality for $W_c$
does not relate to the spatial correlation of random potential,
but simply stems from the tight-binding approximation adopted in all the three models.
Further consideration is needed for clarifying this point.

\end{thebibliography}
\end{document}